\documentclass[sigconf, nonacm]{acmart}
\usepackage[english]{babel}
\usepackage{blindtext}
\usepackage{enumerate}
\usepackage{soul, color, xcolor}
\usepackage{subfigure}

\usepackage{algorithmic}
\usepackage[ruled,vlined]{algorithm2e}
\usepackage{booktabs}

\def\eg{\emph{e.g.,}\xspace}

\def\ie{\emph{i.e.,}\xspace}

\newcommand{\one}{({\em i})\xspace}
\newcommand{\two}{({\em ii})\xspace}
\newcommand{\three}{({\em iii})\xspace}
\newcommand{\four}{({\em iv})\xspace}
\newcommand{\five}{({\em v})\xspace}
\newcommand{\six}{({\em vi})\xspace}

\newcommand{\pb}[1]{\vspace{0.95ex}\noindent{\bf \em #1}\hspace*{.3em}}

\usepackage{multirow}

\newcommand\vldbdoi{XX.XX/XXX.XX}
\newcommand\vldbpages{XXX-XXX}
\newcommand\vldbvolume{14}
\newcommand\vldbissue{1}
\newcommand\vldbyear{2020}
\newcommand\vldbauthors{\authors}
\newcommand\vldbtitle{\shorttitle} 
\newcommand\vldbavailabilityurl{URL_TO_YOUR_ARTIFACTS}
\newcommand\vldbpagestyle{plain}

\renewcommand\footnotetextcopyrightpermission[1]{} 
\setcopyright{none}
\newtheorem{myprin}{Principle}

\def\eg{\emph{e.g.}\xspace}

\def\ie{\emph{i.e.}\xspace}

\newcommand{\squishlist}{
 \begin{list}{$\bullet$}
  { \setlength{\itemsep}{0pt}
     \setlength{\parsep}{3pt}
     \setlength{\topsep}{3pt}
     \setlength{\partopsep}{0pt}
     \setlength{\leftmargin}{1.5em}
     \setlength{\labelwidth}{1em}
     \setlength{\labelsep}{0.5em} } }

\newcommand{\squishlisttwo}{
 \begin{list}{$\bullet$}
  { \setlength{\itemsep}{0pt}
     \setlength{\parsep}{0pt}
    \setlength{\topsep}{0pt}
    \setlength{\partopsep}{0pt}
    \setlength{\leftmargin}{2em}
    \setlength{\labelwidth}{1.5em}
    \setlength{\labelsep}{0.5em} } }

\newcommand{\squishend}{
  \end{list}  }

\begin{document}
\title{Enabling Fast and Flexible Distributed Deep Learning with Programmable Switches}

\author{Heng Pan$^1$, Penglai Cui$^1$, Zhenyu li$^1$, Ru Jia$^1$, Penghao Zhang$^1$, Leilei Zhang$^1$, Ye Yang$^1$, Jiahao Wu$^1$, Jianbo Dong$^1$, Zheng Cao$^2$, Qiang Li$^2$, Hongqiang Harry Liu$^2$, Mathy Laurent$^3$, Gaogang Xie$^4$
}
\affiliation{%
$^1$ICT, CAS, China ~~ $^2$Alibaba Group ~~ $^3$University of Liege ~~ $^4$CNIC, CAS, China
}

\begin{abstract}
Deep learning has been used in a wide range of areas and made a huge breakthrough. With the ever-increasing model size and training data volume, distributed deep learning emerges which utilizes a cluster to train a model in parallel. Unfortunately, the performance is often far from linear speed-up due to the communication overhead between cluster nodes. 

To address this challenge, this paper designs and implements Libra, a network aggregator, that utilizes in-network computation to optimize the communication for distributed DL training in two aspects:
i) reduce active connections and ii) aggregate exchanged network packets. We implemented our Libra on Intel Tofino switches, customized a lightweight host stack and integrated it into an open-source training framework PS-lite. The experimental result shows that our Libra can achieve 1.5$\sim$4$\times$ speedup.  
\end{abstract}

\maketitle

\pagestyle{\vldbpagestyle}

\section{Introduction}
High-dimensional \emph{sparse} data widely exits in Internet-scale deep learning (DL) applications, such as search engine, recommendation systems and online advertising~\cite{jiang2019xdl,10.1145/3503221.3508399}. To enable efficient DL training with sparse data, sparse DL models~\cite{he2017neural,jiang2019xdl,huang2013learning} typically use a two-tier architecture, where the first tier is a \emph{SparseNet} that embeds high-dimensional sparse data into a low-dimensional space via representation learning, and the second is a \emph{DenseNet} that models the relationship between between the dense embedding representation and supervised labels. Two unique features distinguish the sparse DL models from the dense DL ones: \one \emph{data sparsity} that the samples from training data contain a large number of features, but only a few are non-zero; \two \emph{model sparsity} that the most gradients are zero in each training iteration. 

We have witnessed the huge increasing of the size of sparse DL models and the training datasets in recent years. For example, for a business advertising system~\cite{jiang2019xdl}, petabytes (PB) of training data is generated everyday, and the trained DL model consists of billions of features. Indeed, training a sparse DL model is a time-consuming task, and the distributed sparse DL training, which leverages a cluster of nodes to perform training tasks cooperatively, emerges and becomes a practice~\cite{zhang2015optimizing,reagen2016minerva}.

A popular architecture for distributed DL is the parameter server (PS) architecture with data parallelism~\cite{li2014scaling, tiresias, jiang2019xdl}. The training dataset is divided into equal-sized parts (called chunks). In each iteration, workers pull the up-to-date model from parameter servers and perform local training with a chunk of data. The local training results are then sent to parameter servers for the global DL model update. Distributed sparse DL training also follows the above procedure. It has been found that with the increasing of used workers, the communication between workers and parameter servers will become the performance bottleneck~\cite{alistarh2017qsgd,hashemi2018tictac}. Specifically, a recent measurement study~\cite{pan2020imc} reveals that the two factors that hurt the performance: \emph{intensive and bursty communications}.

A straightforward way to mitigate intensive communication is to reduce transmission data volume via gradient compression (\eg gradient quantization~\cite{alistarh2017qsgd} and sparse parameter synchronization~\cite{aji2017sparse}). Other solutions aiming at mitigating communication burstiness decouple the dependency between computation and communication~\cite{hashemi2018tictac,2019Priority,peng2019generic} through scheduling. These solutions improved the performance at the end node side (\ie workers). Another recently emerging direction is to explore the computation capacity in network devices, especially programmable switches, for gradient aggregation~\cite{sapio2017network,sapio2019scaling,switchml,wu2021atp}, which can further accelerate the training process. Nevertheless, \emph{all} the previous in-network aggregation solutions are targeted for \emph{dense} distributed DL models. We found they indeed fall short in accelerating the distributed sparse DL training, because their streaming-based aggregation assumes synchronising \emph{all} the local updates (\ie gradients) in every worker in each training iteration, regardless of whether individual updates are zero or non-zero. This assumption no longer holds in distributed sparse DL training, as only the non-zero embedding vectors and non-zero gradients are transmitted in <key, value> pairs.

To address the above gap, we design and implement \emph{Libra} to enable in-network gradient aggregation for distributed sparse DL training. Libra is built on our key observation from industrial sparse DL applications that the update frequencies of parameters in sparse deep models are extremely biased, where about 50\% of updates are for only the top 30,000 parameters (out of over millions of parameters) (\S~\ref{sec:hot-cold}). Libra thus aggregates the gradients on programmable switches only for these hot parameters, and let the servers aggregate the gradients for the remaining cold ones. That said, rather than offloading the gradient aggregation task for \emph{all} the parameters in~\cite{sapio2017network,sapio2019scaling,switchml,wu2021atp}, Libra only offloads the task for the hot parameters, while keeping the aggregation task for the cold parameters in PS servers.

We address several challenges to implement Libra. First, we propose a sampling-based mechanism to identify the hot parameters by running the training task with a small sample ($4\% - 8\%$) of the whole dataset (\S~\ref{sec:hotidentify}). Second, we design a heat-based parameter placement mechanism at the switch side and a parameter layout (on switch registers) aware gradient packaging mechanism at the worker side to cooperatively reduce the probability that the associated parameters of one gradient packet belong to one register (\S~\ref{sec:para_orche}). In doing so, we reduce the chances that one packet writes a register multiple times in a pipeline, which is not supported by programmable switches. Third, we propose a table-lookup mechanism that enables the on-the-fly floating-point summation on switches, which is essential for gradients aggregation (\S~\ref{sec:approximate}). Last but not least, we enhance the reliability of Libra from the perspectives of packet loss recovery and switch failover (\S~\ref{sec:realiablity}).

We implement Libra and integrated it with PSLite~\cite{pslite} and Intel Tofino~\cite{tofino} programmable switches (\S~\ref{sec:imple}). We perform extensive experiments using a benchmark that includes various sparse DL training tasks (\S~\ref{sec:evaluation}). The results demonstrate the superior performance of Libra, in comparison with the state-of-the-art solutions.  

In summary, our contributions are three-fold: 

\squishlist
    
	\item We design Libra that accelerates the distributed sparse DL training with in-network gradient aggregation on programmable switches. Specifically, it offloads the aggregation task for ``hot'' parameters from PS servers to programmable switches. 
	
	\item We propose the solutions that include the sampling-based hot parameter identification, the heat-based parameter placement on switches, the parameter layout aware gradient packaging on workers, the table-lookup mechanism for the on-the-fly floating-point operation on switches, and two enhancements for improving reliability.

	\item We implement Libra, and integrate it with PS-lite and Intel Tofino switches. Extensive experiments with real-world sparse DL applications demonstrate that Libra improves the aggregation throughput by 1.5$\sim$4$\times$ with limited extra overhead. 
	
\squishend

The remainder of this paper is organized as follows. Section~\ref{sec:background} describes the background and the motivation. We present the design of Libra in Section~\ref{sec:design}. Section~\ref{sec:imple} shows some implementation details. We evaluate our Libra in Section~\ref{sec:evaluation} and present some discussion in Section~\ref{sec:discussion}. We finally discuss the related work in Section~\ref{sec:related} and conclude our work in Section~\ref{sec:conclusion}.

\section{Background and Motivation}\label{sec:background}
\subsection{Distributed Deep Learning}\label{sec:ddl}
\begin{figure}[!ht]
   \centering
   \includegraphics[scale=0.9]{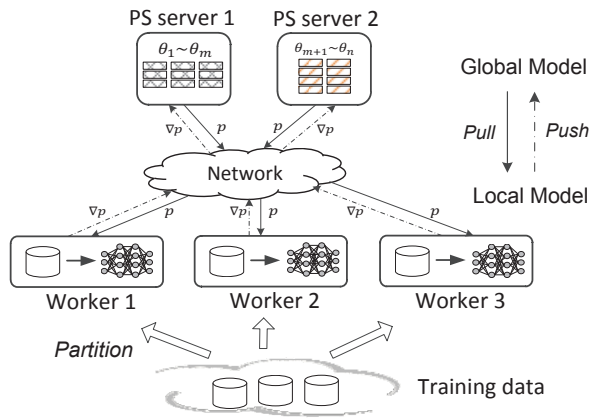}
   \caption{The parameter server architecture.}\label{fig:ps}
\end{figure}

Distributed deep learning (DDL) leverages a cluster of training nodes (called \emph{workers}) to cooperatively train DL models. We consider the widely used \emph{data parallelism} mode, where the training dataset is divided into equal-sized parts to feed training nodes for local training. A widely adopted DDL architecture in industry is the parameter server (PS) architecture (see Figure~\ref{fig:ps}); Microsoft Multiverso~\cite{multiverso}, Alibaba XDL~\cite{jiang2019xdl} and ByteDance BytePS~\cite{peng2019generic} are all built with this architecture. In the PS architecture, workers train their local DL models, while PS servers manage globally shared but non-overlapped DL model parameters. That said, the parameters that a PS server manages are non-overlapped with the parameters of any other PS servers. 
In each iteration of training, workers first pull the up-to-date model from parameter servers, and then perform forward-backward computation with one chunk of training data locally. At the end of an iteration, workers push the trained results (\ie gradients) to the servers for updating the DL model. 

\noindent
\textbf{Synchronous \emph{vs} Asynchronous.} 
Training jobs can be scheduled in either \emph{synchronous} or \emph{asynchronous} mode.
Synchronous training requires PS servers to collect all the gradients from all workers at each integration; it is less effective when workers are equipped with different computation capacities. Asynchronous training, on the other hand, allows workers to work at their own pace, without waiting for other workers to finish their training.

\noindent
\textbf{Reliable Transmission.} Although distributed DL training can tolerate some packet losses due to their special algorithm properties (\eg bounded-loss tolerant)~\cite{xia2019rethinking}, reliable transmission is still a \emph{must} in industry for two reasons~\cite{peng2019generic}. First, distributed DL application developers assume reliable transmission in network substrate; they may optimize their training algorithms based on this assumption. Second, the gradient loss (during transmission) will slow the training convergence and degrade the the end-to-end job performance. As such, the approaches that do not provide reliable transmission may not be viable in practice.

\subsection{Sparse Deep Learning}\label{sec:sddl}
\begin{figure}[!ht]
   \centering
   \includegraphics[scale=0.62]{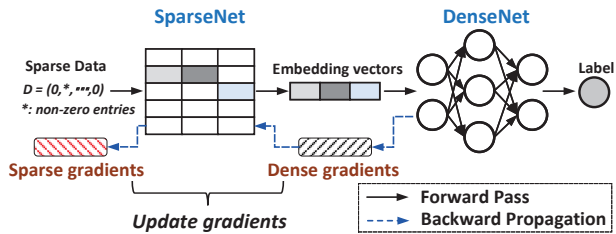}
   \caption{A conceptual architecture of sparse deep training, and it consists of two types of training networks.}\label{fig:sparse_DL}
\end{figure}

High-dimensional sparse data widely exists in Internet-scale applications (\eg search engine and online advertising). Data sparsity could cause low training efficiency if not handled properly; to this end, several sparse DL models have been proposed~\cite{he2017neural,jiang2019xdl,huang2013learning}. These models typically follow a two-tier architecture (see Figure~\ref{fig:sparse_DL}): representation learning (SparseNet) and function fitting (DenseNet). The representation learning embeds high-dimensional sparse data into a low-dimensional space via embedding layers,  while the function fitting models the relationship between dense embedding representation and supervised labels.

\noindent
\textbf{Sparse model training.} Specifically, in the forward pass, a training node reads a batch of sparse data samples, maps them into dense embedding vectors via SparseNet\footnote{Usually, one can look up a huge dictionary or utilize a few CNN/RNN models to implement data mapping.}, and then feeds the results into DenseNet. The backward propagation reverses this data flow and output gradients. The generated gradients in each iteration consist of two parts: the \emph{sparse gradients} for SparseNet and the \emph{dense gradients} for DenseNet; only a few vectors of the SparseNet gradients are non-zero.

\begin{table}[ht]
\centering
\small{
\caption{Neural Network Characteristics of Multiple layer perception~\cite{huang2013learning} (MLP), Crossmedia~\cite{ge2017image} (CM) and Deep interest network~\cite{zhou2018deep} (DIN).}
\begin{tabular}{|c|c|c|}
\hline
\textbf{Deep Model} & \textbf{Neural Net.} & \textbf{\# parameters} \\
\hline
\multirow{2}*{MLP} & SparseNet & 18 Billion\\\cline{2-3}
& DenseNet & 1.2 Million \\
\hline
\multirow{2}*{CM} & SparseNet & 5 Billion\\\cline{2-3}
& DenseNet & 1 Million \\
\hline
\multirow{2}*{DIN} & SparseNet & 18 Billion\\\cline{2-3}
& DenseNet & 1.7 Million \\
\hline
\end{tabular}
\label{tab:sparseNet}
}
\end{table}

Two unique features of the SparseNet distinguish sparse deep training from dense deep training. First, as listed in Table~\ref{tab:sparseNet} which shows the model characteristics of three popular sparse models, the SparseNet uses a much larger training network with billions of parameters than that in DenseNet (millions of parameters). Second, while many gradients of SparseNet in each iteration are zero (\ie sparse), there are still many heavy non-zero vectors. Let us take a NCF~\cite{he2017neural} model with a typical training dataset~\cite{harper2015movielens} as an example. In each iteration, out of 680MB of SparseNet gradients, $\sim$104MB are for non-zero vectors; in comparison, the DenseNet only generates 0.4MB of gradients. 

\noindent
\textbf{Distributed sparse training.} In distributed sparse training, each worker runs the local training job showed in Figure~\ref{fig:sparse_DL}, and synchronizes with other workers for both the SparseNet model and DenseNet model. To reduce the amount of gradient data for transimission, individual workers encode gradient vectors as a list of key-value pairs, and push only the non-zero vectors to PS servers in each iteration. Consequently, the parameters (indices) involved in the transmitted gradients from different workers may not be overlapped. 

\begin{figure}[!ht]
   \centering
   \includegraphics[scale=0.6]{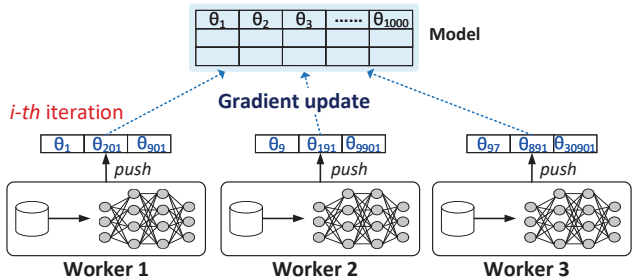}
   \caption{Gradient updates in distributed sparse training.}\label{fig:dsl}
\end{figure}

Figure~\ref{fig:dsl} shows such an example. In the \emph{i-th} iteration, \emph{Worker} 1 pushes 3 non-zero gradients ($\theta_1$, $\theta_{201}$ and $\theta_{901}$) for updating the global model, while \emph{Worker} 2 and \emph{Worker} 3 transmit non-zero gradients for other parameters (\eg \{$\theta_9$, $\theta_{191}$, $\theta_{9901}$\} from \emph{Worker} 2). It is worth noting that the parameters with non-zero gradients in individual workers are unknown in advance.

\subsection{Limitations of Prior In-network Aggregation}\label{sec:limitation}
Recently, a large amount of research effort~\cite{switchml,wu2021atp,sapio2019scaling} has been devoted to \emph{in-network gradient aggregation} in order to reduce the communication volume and improve the overall training performance. Specifically, programmable switches are exploited to sum the gradients sent by workers and send only the summation results to PS servers. By doing so, the data volume to PS servers is greatly reduced.

We particularly focus on two state-of-the-art systems that leverage programmable switches for in-network gradient aggregation: SwitchML~\cite{sapio2019scaling} and ATP~\cite{wu2021atp}. 
While being effective for dense training, they fall short in speeding up \textbf{sparse} DL training because of their streaming-based aggregation.

\begin{figure}[!ht]
   \centering
   \includegraphics[scale=0.55]{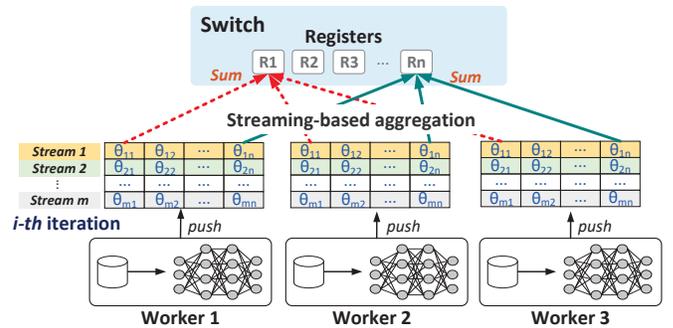}
   \caption{The Streaming-based in-network gradient aggregation in~\cite{sapio2019scaling, wu2021atp}.}\label{fig:dense}
\end{figure}

To show this, we briefly describe their workflow. As illustrate in Figure~\ref{fig:dense}, the gradients are chunked into $m$ streams at each worker, where the $j$-th ($j<m$) stream in each worker contains the gradients for the same set of parameters. Each worker sends one stream at each time slot to programmable switches for aggregation. Programmable switches will aggregate (\ie sum) the gradients of the $j$-th streams from all workers. This approach works well in dense deep learning, where in each iteration, the gradients of all the parameters need to be sent out for aggregation. Besides, because of the limited storage space in programmable switches, some approaches (\eg ATP) require synchronization of time slots among workers --- all workers not to transmit the
next stream of gradients until the programmable switches relieve the last aggregated results. That said, they may not support asynchronous training. 

The above streaming-based approach does not work for deep sparse training, simply because in each iteration, only the non-zero gradients are sent to PS servers for aggregation. Worse still, the parameters with non-zero gradients in individual workers are unknown in advance. Simply applying existing approaches will lead to limited chances of aggregations on programmable switches, or even incorrect aggregation.

Another design component in existing approaches that we aim to improve is about floating-point arithmetic. Because the state-of-the-art Tofino programmable switches only support integer summation~\cite{Yuan.NSDI22}, both SwitchML and ATP adopt a float-to-integer approach that converts floating-point gradients into integers via multiplying a scaling factor at the worker side. To reduce the accuracy loss introduced by this float-to-integer approach, a negotiation mechanism may be used. For instance, SwitchML requires all workers to negotiate with each other to decide an appropriate value of the scaling factor before transmitting the streams of gradient. Such frequent negotiations would introduce performance overhead. Another possible solution is to equip each switch with a floating-point unit (using FPGAs or ASICs) directly as in~\cite{yuan2022unlocking}. Unfortunately, this solution indeed increases the cost of switches while those ``legacy'' programmable switches that have been in widespread use today often do not have such units. That said, our Libra aims to calculate float-point gradients on the ``legacy'' programmable switches directly with high accuracy and low overhead.

\section{Libra Design}\label{sec:design}
This section presents the design of Libra that is able to accelerate \textbf{asynchronous} distributed \textbf{sparse} DL training. We begin with the key observation that encourages the design of Libra, and then detail Libra's main components.

\subsection{Characterizing ``Hot-cold'' Phenomenon}\label{sec:hot-cold}
We first present our observation on the highly skewed update frequency of parameters in distributed sparse DL training. This observation is derived from our (industrial) training tasks of two typical sparse deep learning models: search engine and online advertising recommendation.

\pb{Experimental setup.} We use a testbed with 16 workers and 1 PS servers to train DDL models for online advertising recommendation (Task 1) and search engine (Task 2). The model of Task 1 consists of 150 million parameters, while Task 2 contains about 9 million model parameters. During training, we collect logs from the PS server, and count the update frequency of each parameter. 

\begin{figure}[!ht]
   \centering
   \begin{minipage}{0.7\linewidth}
   \centerline{\includegraphics[width=\linewidth]{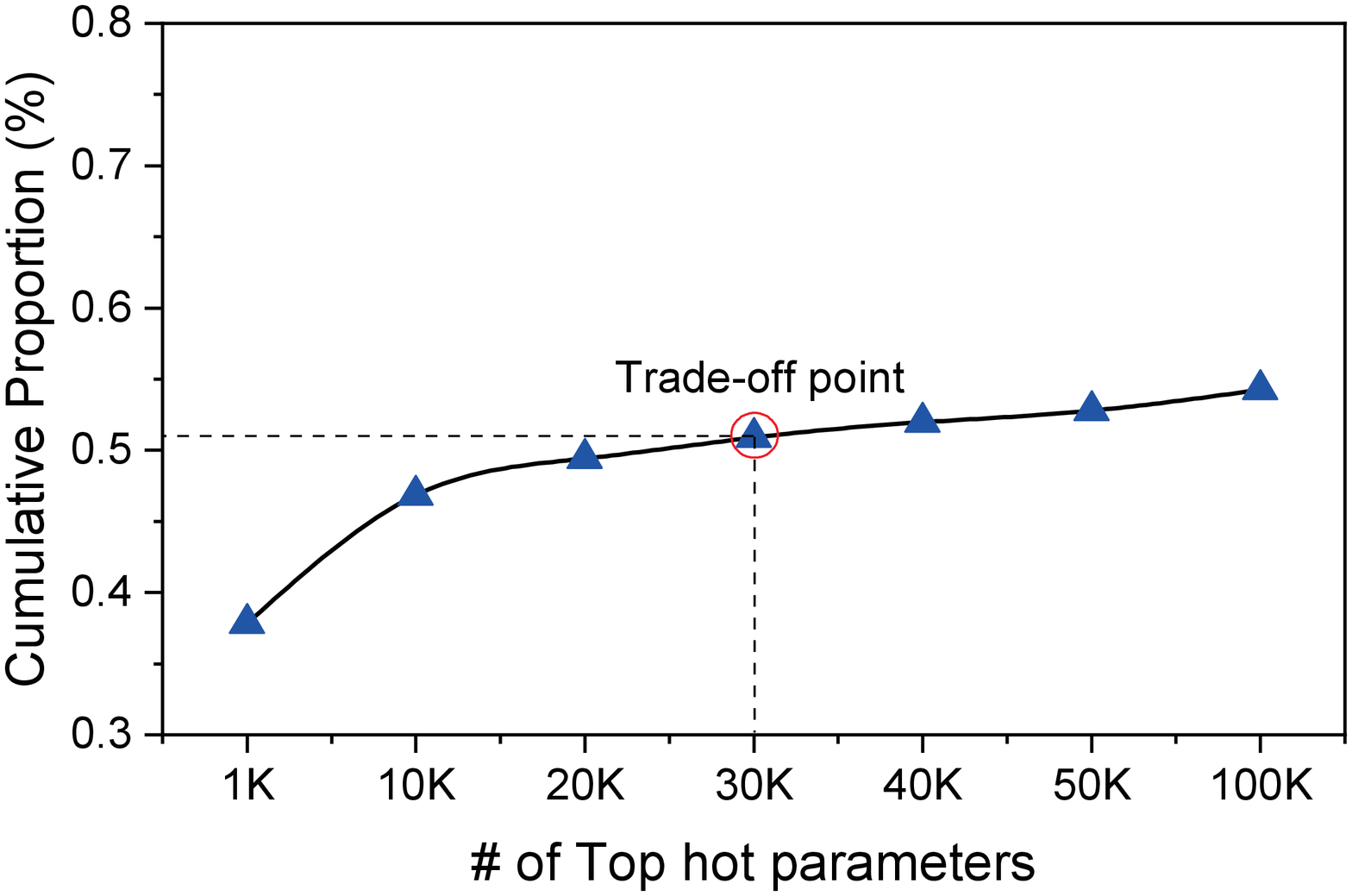}}
   \centerline{(a) Recommendation}
   \end{minipage}
   \vfill
   \begin{minipage}{0.7\linewidth}
   \centerline{\includegraphics[width=\linewidth]{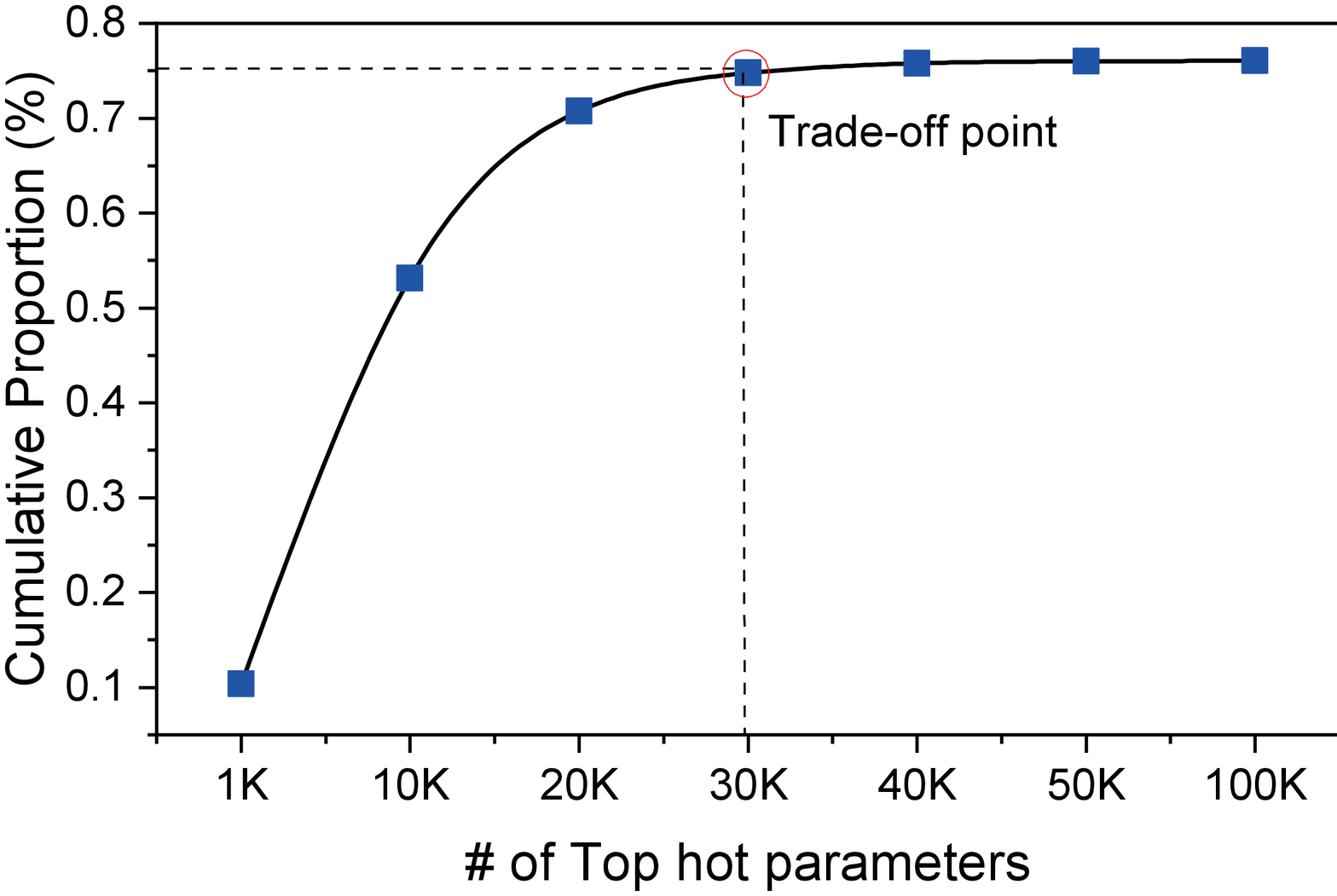}}
   \centerline{(b) Search}
   \end{minipage}
   \caption{Cumulative distribution of the parameter update frequency for two production sparse models.}\label{fig:hotpar}
\end{figure}

\pb{Observation.} We sort the DL model parameters based on their update frequency, and present the cumulative proportion of update frequency for the top 100K parameters in  Figure~\ref{fig:hotpar}. Our most surprising finding is that they contain very ``hot'' parameters who contribute the bulk of the update. For example, across the 150 million parameters, the top 30,000 parameters of Task 1 constitute over half of all updates. Likewise, for Task 2, its top 30,000 parameters even account for about 70\% of updates.

\pb{Summary.} High-dimensional sparse deep learning tasks exhibit a \emph{hot-cold phenomenon} where the hot parameters contribute most traffic. Some independent research ~\cite{pan2020imc} also reported this phenomenon. Putting this in the context of in-network gradient aggregation, we envision a system that aggregates the gradients only for the ``hot'' parameters as they contribute most of the updates. By doing so, we can accelerate the distributed sparse training with a limited storage requirement on programmable switches.

\subsection{Design overview}
\label{sec:aggregation}

\begin{figure}[!ht]
   \centering
   \includegraphics[scale=0.83]{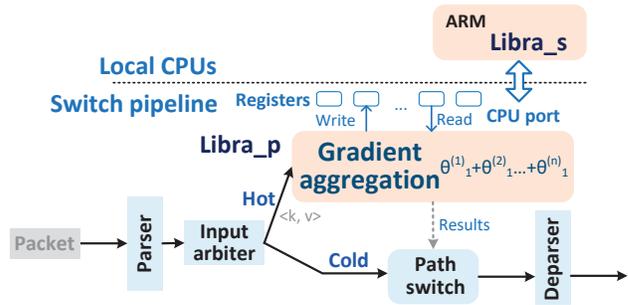}
   \caption{\small{The overview of Libra.}}\label{fig:sparse_agg}
\end{figure}

To restate, our Libra uses programmable switches for in-network sparse gradient aggregation. Figure~\ref{fig:sparse_agg} shows its two components on the switch, namely $Libra\_p$ and $Libra\_s$. $Libra\_p$ is deployed on the switch pipeline to implement gradient aggregation, while $Libra\_s$ runs on the local CPUs of the switch to facilitate reliable transmission (see Section~\ref{sec:realiablity}). Given the ``hot-cold'' phenomenon we observed, Libra adopts a hot-cold separation mechanism. Specifically, it selects those ``hot'' model parameters that are frequently updated for aggregation on switches, while the remaining ``cold'' parameters are handed over to the parameter servers without in-network aggregation. It is noteworthy that such an separation does not require  synchronization between workers; workers can pull the up-to-date aggregation results from the switch and the parameter servers.

\subsection{Identifying Hot Parameters} \label{sec:hotidentify}
Without running the training work, the hot parameters are unknown. In fact, the hot parameters cannot be accurately predicted because the update frequency of parameters do not follow a pre-defined pattern and thus may vary from a sparse DL task to another (see Figure~\ref{fig:hotpar}).

To this end, we adopt a sampling-based mechanism to approximately capture the update frequency distribution of DL model parameters. To this end, we first extract a small training dataset by randomly sampling the whole dataset. We then train the DL model with the sampled  dataset and record the update frequency of each model parameter. Finally, we sort the parameters based on their update frequency, and label the top $k$ parameters as hot ones. As we show in \S\ref{sec:hot-precision}, by running the training with only 8\% of the whole dataset as input, we can identify hot parameters with an accuracy 90\%. 

The number of hot parameters ($k$) whose gradients will be aggregated also depends on the available resources on switches. Specifically, we identify hot model parameters based on Principle~\ref{def:hot}.

\begin{myprin}\label{def:hot}
(Hot parameter Identification). Consider a list of parameters $\theta$=\{$\theta_1$, $\theta_2$, ..., $\theta_n$\} that are ranked in descending order by the update frequency. We use $UF$=\{$uf_1$, $uf_2$, ..., $uf_n$\} to refer to their corresponding update frequency. We also assume that one programmable switch is equipped with 20MB on-chip memory, and storing a model parameter in a switch consumes 4 bytes of memory. We say that the top-$k$ parameters (\{$\theta_1$, ..., $\theta_k$\}) are hot if they satisfy the following two conditions: 
$$
\left\{
\begin{array}{l}
T_k/T_n \geq p \\
4B\times k \leq c \times 20MB
\end{array}
\right.
$$
where $T_k$=$\sum_{i=1}^kuf_i$, $T_n$=$\sum_{i=1}^nuf_i$, $p\in (0, 1)$ and $c\in (0, 1)$ are two design parameters.
\end{myprin}

The parameter $c$ is the fraction of on-chip memory that we would like to use for gradient aggregation. In practice, $c$ should be small ($0.05\sim 0.1$) because occupying too much memory would affect conventional functions of switches (\eg packet forwarding). The parameter $p$ is the expected proportion of traffic that will be intercepted and processed by the switch. While a larger $p$ is preferred (so as $k$), more memory would be required. Nevertheless, as showed in Figure~\ref{fig:hotpar}, the marginal benefit in terms of traffic saving (\ie the growth of $p$) when increasing $k$ beyond some points (called trade-off points) becomes very small. Let us take Figure~\ref{fig:hotpar}(b) as an example. Increasing $k$ beyond $30K$ will bring very limited benefit in terms of traffic saving. In this example, $k$ will be set as 30,000 in our implementation. 

\begin{figure}[!ht]
   \centering
   \includegraphics[scale=0.8]{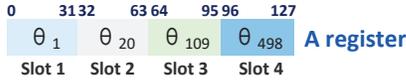}
   \caption{\small{An example of a register caching multiple parameters.}}\label{fig:regsiter}
\end{figure}

\subsection{Parameter Orchestration} \label{sec:para_orche}
\pb{Parameter layout in switch registers.} The on-chip memory in a switch is physically organized as registers. A register is similar to an array (see Figure~\ref{fig:regsiter}), which consists of multiple register slots. One hot parameter takes one register slot to cache its gradient; a new gradient will be added to the cached gradient to get the final summation of all the gradients of this parameter. A practical restriction on switches is that one register can be operated only once in one pipeline. That said, if a register cached gradients of two parameters whose updates are carried in one packet, they would not be able to be aggregated in one pipeline. While we could use the \emph{recirculation} operation in programmable switches to  recirculate the packet back to the pipeline, recirculations can degrade the performance. As such, we need to carefully assign hot parameters to registers so that the chances that the updates of two parameters assigned in one register are carried in the same packet is low. 

Before delving into the detail of parameter layout, we first describe the mapping of parameter indices. Gradients in distributed deep sparse training are transmitted in the form of $<key, value>$ pairs, where $key$ is the index of a parameter. Because we will use the index to directly locate their corresponding register slots, we need to map the indices obtained from the models into the range of $[0, M-1]$, where $M$ is the number of register slots used for gradient aggregation. Suppose we rank the parameters in descending order based on the update frequency, then a parameter's index after mapping is its rank (from 0 to $M-1$). Workers store the mapping information locally, which is used to restore the indices of the aggregation results.

\begin{figure}[!ht]
   \centering
   \includegraphics[scale=0.9]{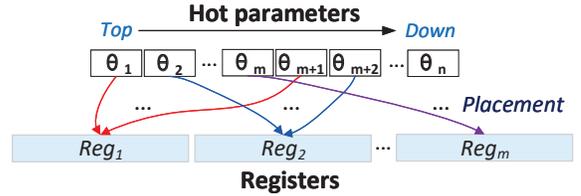}
   \caption{\small{The heat-based parameter placement, where $m$ registers are used to cache $n$ hot parameters.}}\label{fig:aggregation}
\end{figure}

Next we describe the parameter layout in registers using a heat-based parameter placement mechanism (see Figure~\ref{fig:aggregation}). Specifically, let us consider a list of hot parameters \{$\theta_{1}$, $\theta_{2}$, ..., $\theta_{n}$\} ranked in descending order of their update frequency (\ie \emph{heat}). Our method places the $n$ parameters into $m$ registers as follows. The $i$-th register, stores the (${i+m*j}$)-th parameters, where $0 \leq j\leq \lfloor n/m\rfloor$. Our intuition is that the ``heat'' distribution of the hot parameters is non-uniform (see Figure~\ref{fig:hotpar}), so the heat of $\theta_i$ would be much higher than that of $\theta_{i+m}$. Therefore, the probability that the gradients of $\theta_i$ and $\theta_{i+m}$ are encapsulated into one packet is low. That said, we can use one register to store $\theta_i$ and $\theta_{i+m}$. With this basis, we deduce the above conclusion that $\theta_{i+m*j}$ where $0 \leq j\leq \lfloor n/m\rfloor$ are assigned to one register. We will show the effectiveness of this layout in \S\ref{sec:exp_aggr}.

\pb{Packaging gradients at workers}. 
At the worker side, we adopt a parameter layout aware gradient packaging mechanism for encapsulate gradients into packets (see Algorithm~\ref{algo:encapsulated}). The core idea is to encapsulate parameter gradients into packets in order to achieve two goals: \one reducing the likelihood that the parameters encapsulated to one packet are cached in one register; \two using as few packets as possible. We assume that workers know the number of registers $m$. When a batch of gradients for the hot parameters need to be transmitted, we use Algorithm~\ref{algo:encapsulated} for packaging gradients into packets. The algorithm first estimates how many packets\footnote{Because we need to parse the whole packet by the parser of programmable switches, the packet size is limited to 192 bytes.} (denoted by $P$) are needed to carry these parameter gradients (line 2), and then process each parameter $\theta$ as follows, where $\theta$ is the (rank) index of the parameter.

\begin{itemize}
\item First, it uses the index of $\theta$ to locate the register ID ($k$) that caches $\theta$ (line 4) at switches;

\item Second, it filters out the packets that have carried at least one parameter that also belongs to the $k$-th register (line 7 to 9); 

\item Third, it appends $\theta$ in a candidate packet and updates corresponding states. (line 10 to 16). 

\item Forth, if there is no such packet, $\theta$ will be recorded to a set ($G^{'}$) for further processing (line 17 to 18).
\end{itemize} 

\begin{algorithm}[!ht]
\DontPrintSemicolon
\LinesNumbered
\KwIn{$G$, a batch of gradients to be transmitted.}
\KwIn{$m$, the number of switch registers.}
\KwOut{return some packets that carry the gradients.}
$P\gets$ \FuncSty{Estimate\_packets}\ArgSty{(G)};\;
$G^{'}\gets \emptyset$; $P_1\gets P$; \tcp{P is a list of packets.}
\tcp{Packaging G to $n$ packets}
\For {$\theta \in G$} {
	$k\gets\theta.id\%m$; \tcp{Get register ID} 
	$P_2\gets P_1$; \tcp{$P_2$ carries candidates for $\theta$}
	$is\_inserted\gets false$;\;
	\tcp{Find out pkts having already include the parameters sharing the same register with $\theta$}
	$res\gets$ \FuncSty{same\_reg\_pkt\_find}\ArgSty{(k)}; \;
	\If {$res \not= NULL$} {
		$P_2.$\FuncSty{erase}\ArgSty{($res.$\FuncSty{begin}(), $res.$\FuncSty{end}())};\;
	}
	$pkt \gets P_2.$\FuncSty{first}\ArgSty{()};\;
	\If {$pkt \not=$ NULL} {
		\tcp{$pkt$ will carry $\theta$.}
		\tcp{$P_1$ tracks the full state of pkts.}
		\FuncSty{Update}\ArgSty{($P_1$, $pkt$, $\theta$)};\;
		\FuncSty{Update}\ArgSty{($P$, $pkt$, $\theta$)}; \tcp{Final results.}
		$is\_inserted\gets true$;\;
		\If {\FuncSty{is\_full}\ArgSty{(pkt) == True}} {
			$P_1.$\FuncSty{erase}\ArgSty{($pkt$)};\;
		}
	}
	\If {$is\_inserted \not= true$} {
		$G^{'}.$\FuncSty{append}\ArgSty{($\theta$)};\;
	}
}
\If {$G^{'} \not= \emptyset$} {
	$P.$\FuncSty{insert}\ArgSty{(\FuncSty{create\_pkt\_padding}\ArgSty{($G^{'}$)})};\;
}
\FuncSty{return} \ArgSty{$P$};\;
\caption{{\sc Parameter\_orchestrating}(\ArgSty{G, m})}
\label{algo:encapsulated}
\end{algorithm} 

Finally, if $G^{'}$ is not empty, the worker will encapsulate the parameter gradients in $G^{'}$ into a number of packets without considering the parameter layout in switch registers (line 19-20). The reason why we do not use the packets in $P$ to carry the parameters in $G^{'}$ is as follows. Let us assume that the number of parameters in $G^{'}$ is $n$. If we used the packets in $P$ to carry $G^{'}$, it would lead to $n$ times recirculation operations. On the contrary, if we encapsulate them into new packets directly, the usage of recirculation operations can be significantly saved, because the parameters in $G^{'}$ are unlikely to belong to one register.

\subsection{Gradient aggregation on switches}\label{sec:approximate}

Switches parse the received gradient packets using their programmable parser. By doing so, the switch will get the a list of <key, value> pairs, each of which is the gradient (value) of a parameter with \emph{key} as the index. Specifically, for an update of the parameter $\theta$, the switch locates the register as well as the slot in the register that caches $\theta$. More specifically, we use a hash table to map the the index of one parameter into the position of the registers. 
Then the update (\ie gradient) is added to the cached value.

\pb{Floating-point summation on switches. } Gradients are typically represented as 32-bit floats. Nevertheless, the pipeline in programmable switches does not support floating-point summation. Rather than using the aforementioned floating-to-integer method that may introduce accuracy loss, we propose a table-lookup method that enables the on-the-fly floating-point summation for 32-bit floats. Our method is inspired by~\cite{cui2021netfc} that implements floating-point operations for 16-bit floats.

For any two positive numbers $x$ and $y$, the following equation holds: $x+y=2^{i+log_2(1+2^{j-i})}$, where $ i=\log_2(x)$ and $j=\log_2(y)$. 
A 32-bit floats in IEEE 754 consists of three portions: the sign field (1 bit), the exponent field (8 bits) and the fraction field (23 bits); they together form a key for table lookup. Thus, a strawman solution is to set up three tables on switches: \one a \emph{logTable} that is used to record the logarithm values of all possible keys; \two a \emph{miTable} that is used to get $\sigma(\theta) = \log_2(1+2^{\theta})$ for a given $\theta$; \three an \emph{expTable} that is to get the exponential value for a given key. Let us consider two 32-bit floats $x$ and $y$ to illustrate the floating-point operations using the these 3 tables. Both $x$ and $y$ are first mapped into the logarithmic number system~\cite{kingsbury1971digital} by looking up the first table (\emph{logTable}), yielding $\log_2(x)$ and $\log_2(y)$. Next it computes $\theta=\log_2(y)$ minus $\log_2(x)$ and uses $\theta$ to look up the second table (\emph{miTable}) and gets $\log_2(1+2^{\log_2(y)-\log_2(x)})$, which then is added by $\log_2(x)$. Finally, the summation is taken as the key to look up the third table (\emph{expTable}) to get the value of $x+y$.

Unfortunately, the bit width of the key in \emph{logTable} can be up to 32 bits, so it would consume $\sim$16GB ($2^{32}\times 4B$) on-chip memory, which is unacceptable in practice. To this end, we further propose an approximate method to divide the large \emph{logTable} into a few small tables in order to reduce the memory consumption. More specifically, a positive IEEE 754 32-bit float $p$ is represented as $p$=$2^{e-127}*1.f_1f_2...f_{23}$, and we assume that $m=1.f_1f_2...f_{11}$ and $\Delta m=0.f_{12}f_{13}...f_{23}*2^{-11}$. Consequently, $p=2^{e-127}*$($m$+$\Delta m$), and $log(p) = (e-127)+log(m+\Delta m)$. For any $log(m+\Delta m)$, it can be converted into a polynomial via Taylor series~\cite{Taylorseries}, and further be approximated as $log(m)+2^{log(\Delta m)-log(m\ln2)}$. Thus, $log(p)$ can be simplified as follow.

\begin{equation}
\begin{aligned}
log(p) &= (e-127)+log(m+\Delta m)\\
&= (e-127)+ log(m)+\frac{1}{m \ln2}\Delta m-\frac{1}{m^2 \ln2}\Delta m^2+\cdot\cdot\cdot \\
&\approx (e-127)+log(m)+\frac{1}{m \ln2}\Delta m \\ 
&=(e-127)+log(m)+2^{log(\Delta m)-log(m \ln2)}
\label{form:32-bit-formulation}
\end{aligned}
\end{equation}

The approximation error is negligible because $\Delta m$ is far less than $m$. Due to space limitation, we omit the related theoretical proof, but confirm the error is very limited through experiments (see \S~\ref{sec:fp}).  

Consequently, the huge \emph{logTable} is replaced by five small tables: an 8-bit \emph{epoTable}, three 12-bit \emph{logTable} and a 16-bit \emph{expTable}. More specifically, we use the \emph{epoTable} to obtain the result of $(e-127)$, which is then used to look up the \emph{logTables} to get $log(m)$, $log(\Delta m)$ and $log(m \ln2)$. Finally, the \emph{expTable} is used to obtain $2^{log(\Delta m)-log(m \ln2)}$. 

\begin{figure}[!ht]
   \centering
   \includegraphics[scale=0.68]{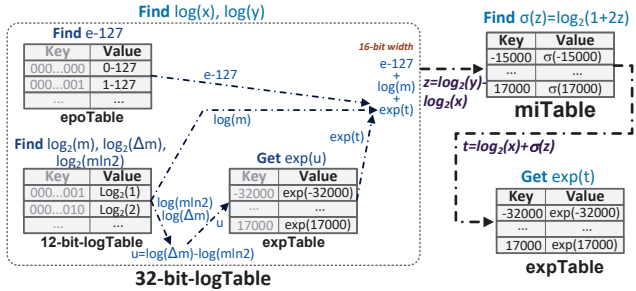}
   \caption{32-bit floating-point arithmetic on switches.}\label{fig:approximate}
\end{figure}

In summary, we need seven tables to implement the float-pointing summation on switches; the total storage demand for these tables is only 408.5KB (=$256*2B+3*4096*2B+65536*2B+65536*2B+65536*2B$)\footnote{While this on-chip memory demand is acceptable in practice, we can also leverage the prefix-based compression proposed in~\cite{cui2021netfc} to further reduce the memory consumption.}. Figure~\ref{fig:approximate} summarizes the procedure for the 32-bit float-point summation on switches using our table-lookup mechanism. 

\subsection{System Reliability}\label{sec:realiablity}
The system reliability is important in production environments. We consider two sources of failures that would undermine Libra: packet loss and equipment failure.

\pb{Packet Retransmission.} While packet loss happens much less frequently in data center networks ($<10^{-3}$~\cite{10.1145/2785956.2787496}), it indeed may happen. A packet loss would result in the loss of all the gradients the packet carries and thus degrading the training performance. To this end, we leverage the per-packet ACK mechanism for loss detection and retransmit the lost packets. Specifically, the receiver of a packet will immediately return an ACK to the sender. That said, switches will ack the gradient packets sent from workers; PS servers will ack the packets for cold parameters; and workers ack the aggregation packets that are sent either from switches or PS servers. The sender of a packet will mark the packet as loss if it does not receive the ack before the timer expires; the retransmited packet has the same sequence number as the original one.

To implement the above mechanism, switches need to keep the unacked aggregation packets locally for possible retransmissions. To this end, when sending out aggregation packets to workers, switches also forward one copy to its local CPUs ($Libra\_s$). That said, the local CPUs are only involved for reliable transmission. 

\begin{figure}[!ht]
   \centering
   \includegraphics[scale=0.75]{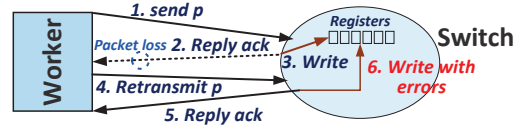}
   \caption{An illustration of the repeat-write-error.}\label{fig:interaction}
\end{figure}

In practice, we also observe the other error related to packet loss: \emph{repeat-write-error}, where the ack packet is lost (see Figure~\ref{fig:interaction}). It happens in the case that the ack packet sent from a switch to a worker is lost, but the switch has aggregated the gradients carried by the packet that is being acknowledged. The worker deems the gradient packet has been lost, so it retransmits the packet. Without a local recording in the switch, the switch will wrongly aggregate the gradients twice. 

We address the issue caused by the {repeat-write-error} as follows. Workers will explicitly mark the retransmitted packet (\eg using one bit in the packet hearder). When receiving a retransmitted packet, the switch will check the local records to figure out whether the gradients in the packet has been aggregated. The local records can be implemented either in local CPUs on the switch or using a Bloom Filter in the pipeline. Our current implement relies on local CPUs.

\pb{Failover mechanism.} Switches may fail unexpectedly in practice. For the traditional switches that only forward packets, we can reroute packets to detour around the failed switches. However, because now the switches have been leveraged for gradient aggregation, the above bypass-based mechanism will result in the loss of aggregated gradients on programmable switches and thus crash down the training task.

\begin{figure}[!ht]
   \centering
   \includegraphics[scale=0.85]{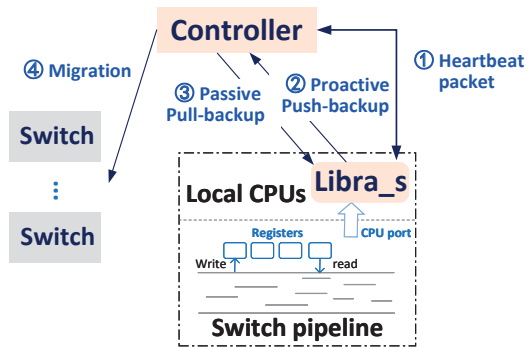}
   \caption{\small{A detection-migration based failover mechanism.}}\label{fig:migration}
\end{figure}

To this end, we design a detection-migration failover mechanism (see Figure~\ref{fig:migration}), which leverages the controller of the switches for the detection of failure and invoking the task migration. Specifically, the controller periodically requests the status of switches through heartbeat messages. Switches reply with statistics of the switches (\eg resource utilisation). The controller then uses these statistics along with the response delay to detect whether the switch is about to fail (\eg high packet loss rates, high on-chip memory usage). If the controller finds out the abnormal status of the switch via the heartbeat packets, the switch states (a.k.a aggregation results) are passively pulled by the controller. 

Because the layout of the hot parameters on registers are pre-defined, Libra can deploy $Libra\_s$ and $Libra\_p$ to the standby switches in advance. The controller selects one of these standby switches and sends the states to this switch. The selected switch then uses the migrated states to resume the training. 

\begin{figure*}[ht]
   \centering
   \begin{minipage}{0.19\linewidth}
   \centerline{\includegraphics[width=\linewidth]{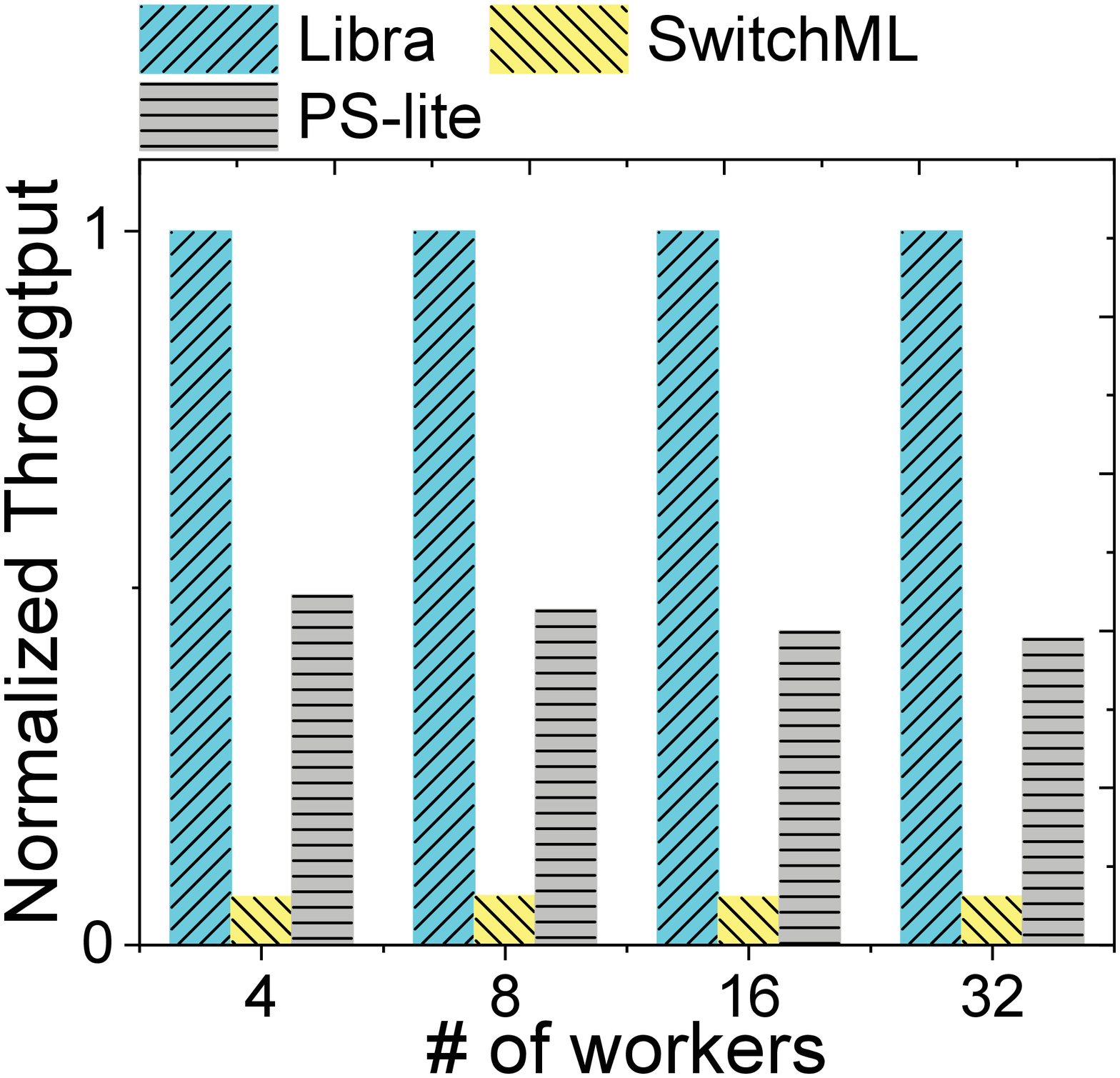}}
   \centerline{\small{(a) DeepLight}}
   \end{minipage}
   \begin{minipage}{0.19\linewidth}
   \centerline{\includegraphics[width=\linewidth]{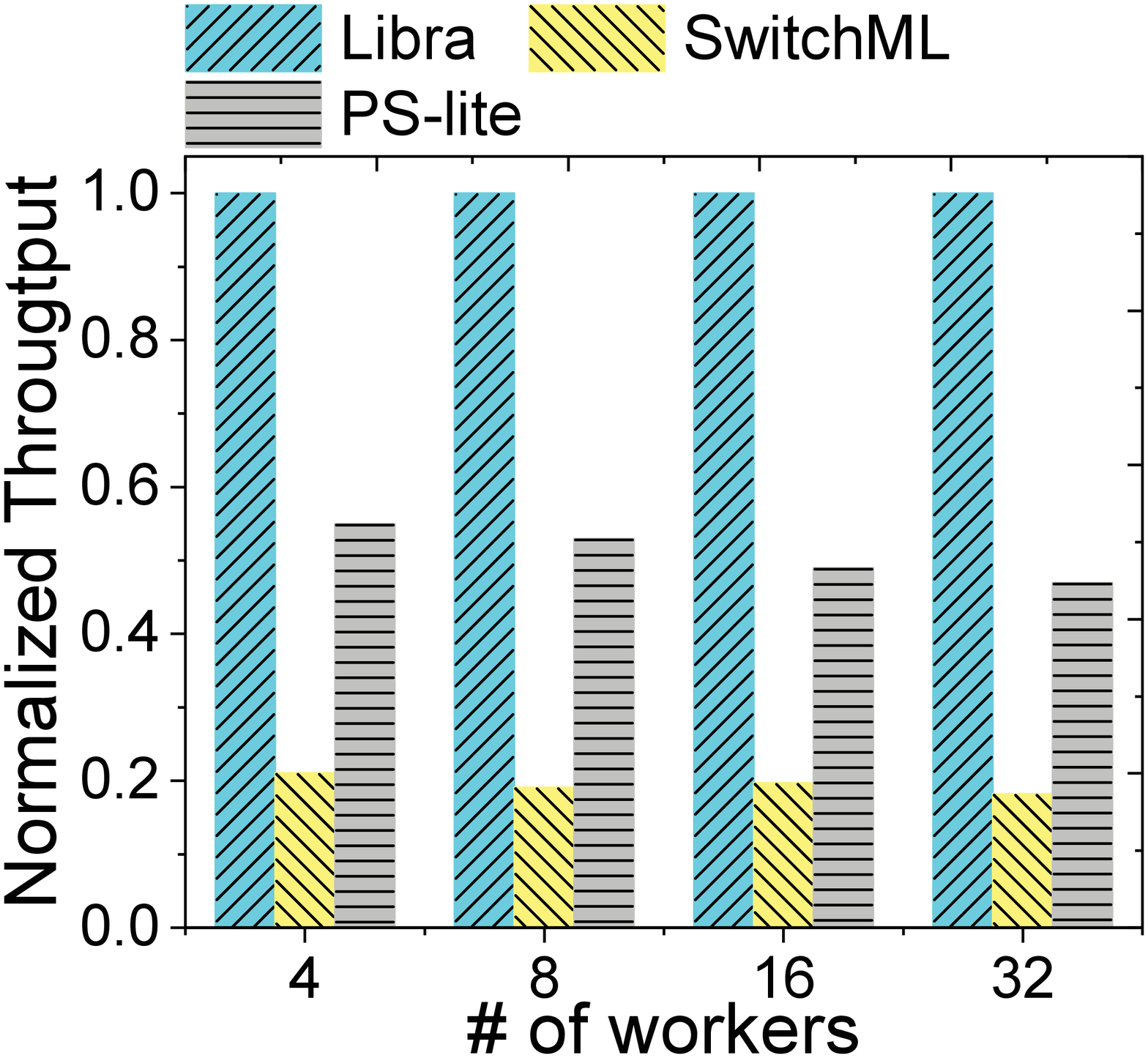}}
   \centerline{\small{(b) LSTM}}
   \end{minipage}
   \begin{minipage}{0.19\linewidth}
   \centerline{\includegraphics[width=\linewidth]{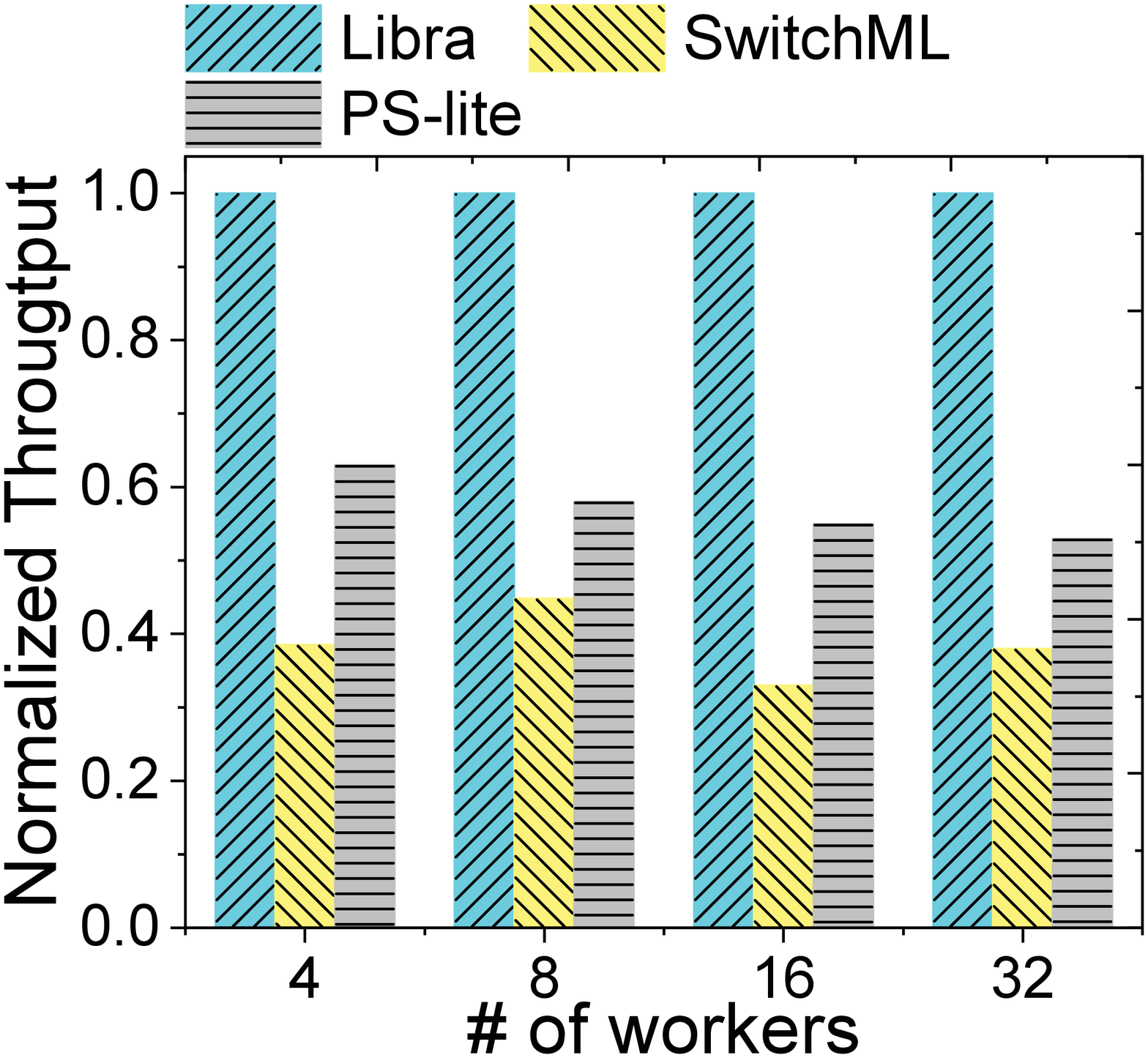}}
   \centerline{\small{(c) NFC}}
   \end{minipage}
   \begin{minipage}{0.19\linewidth}
   \centerline{\includegraphics[width=\linewidth]{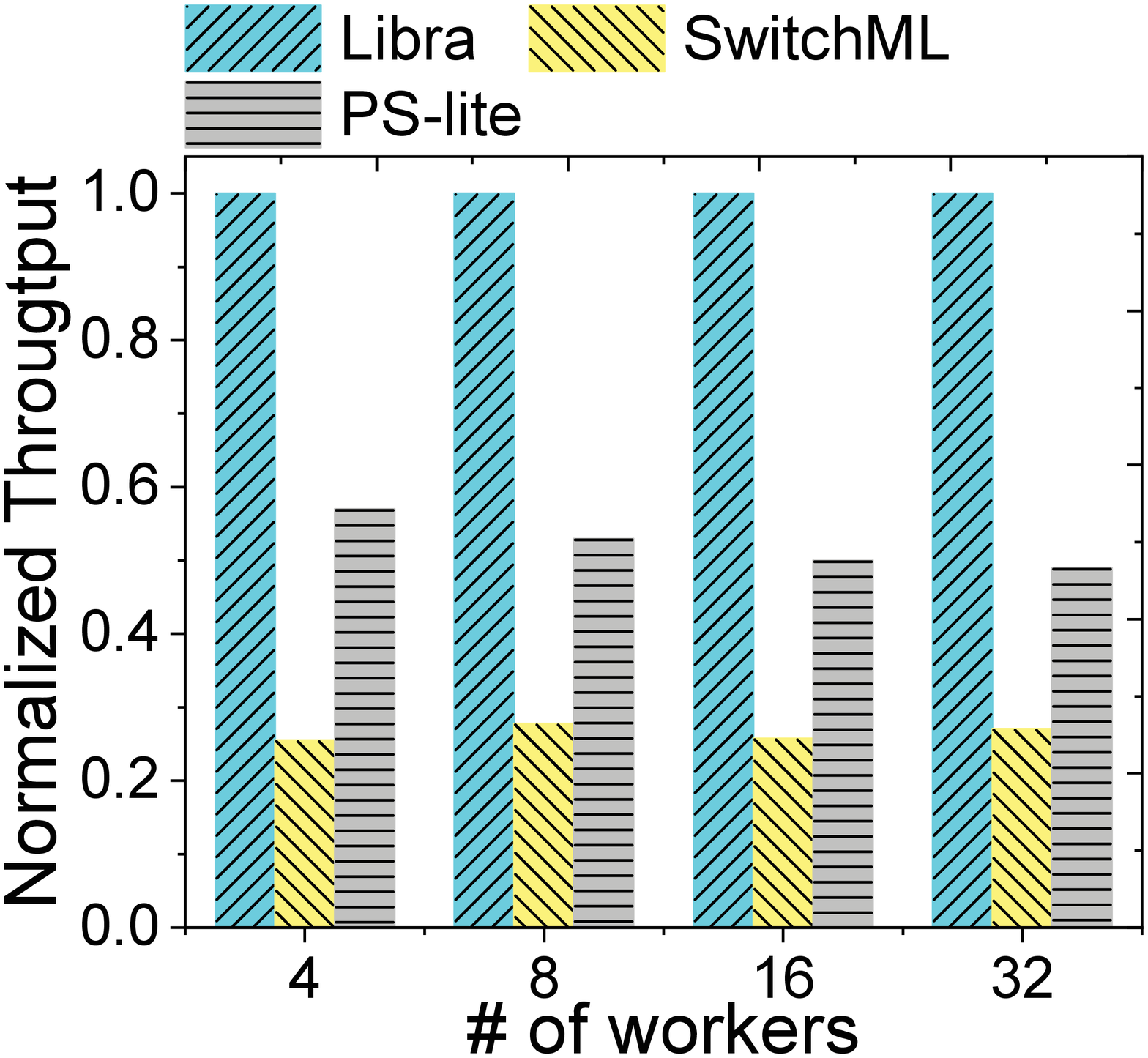}}
   \centerline{\small{(d) OA}}
   \end{minipage}
   \begin{minipage}{0.19\linewidth}
   \centerline{\includegraphics[width=\linewidth]{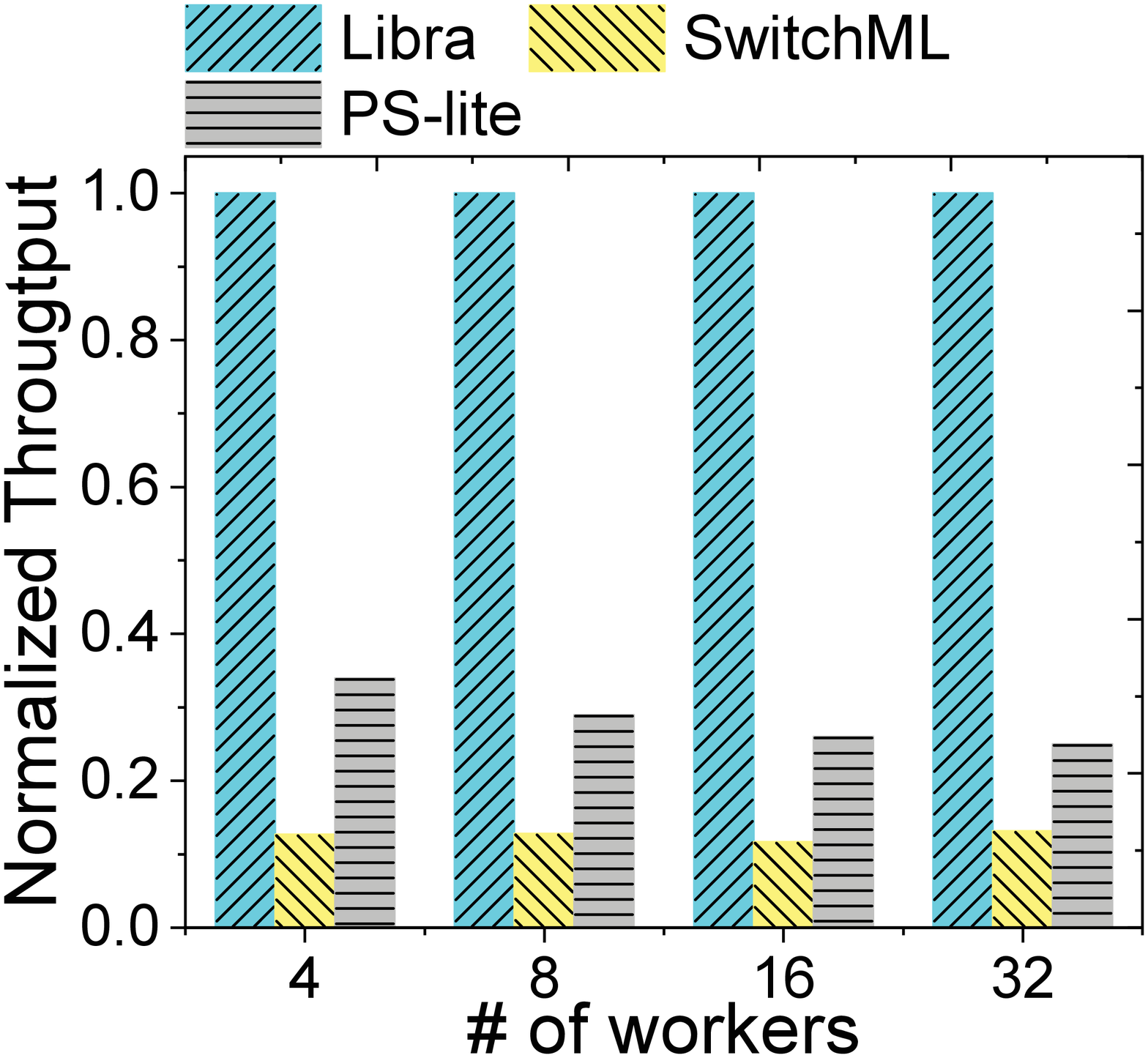}}
   \centerline{\small{(e) SE}}
   \end{minipage}
   \caption{Aggregation throughput normalized by that of Libra.}\label{fig:vs-ps-lite}
\end{figure*}

\section{Implementation}\label{sec:imple}
We implement a prototype of Libra on commodity servers and programmable switches. Specifically, data plane components at switches are implemented with $P4_{16}$ language (1500+ LoC) and compiled with the Intel Capilano software suite; they are deployed to 3.2 Tbps Intel Tofino switches\footnote{Each switch consists of two pipelines (i.e., ingress and egress pipelines), which contains 12 stages in total.}. Libra host stack is customised on lwIP~\cite{lwip}, a lightweight user-level TCP/IP stack. Finally, we further integrate our Libra into PS-lite\cite{pslite}, an open-source parameter server platform.

\noindent
\textbf{Data plane component.} Programmable switches like the Intel Tofino switch that we use have some restrictions on the access of registers and tables. A particular restriction that we have to deal with is that a register can only be read and written once per packet processing. In case we need to read and write a register more than once, we adopt the \emph{recirculation} operation that recirculate to the packet back to the pipeline. In our implementation, a packet is at most recirculated once, \ie it goes through the pipeline twice at most.
 
\noindent
\textbf{Host Stack.} lwIP is a lightweight user-level TCP/IP stack; the current versions running on Linux servers only provide TAP/TUN virtual adapters. TAP/TUN adopters incur multiple packet copy operations and context switches between the kernel space and user space, degrading the whole performance. To this end, we turn to DPDK (Data Plane Development Kit)~\cite{dpdk}. Specifically, we use DPDK APIs to implement a network device driver based on the \emph{template} provided by lwIP; by doing so, lwIP can send/receive packets at line speed.

\noindent
\textbf{System integration.} PS-lite utilizes ZeroMQ~\cite{zmq} (a.k.a zmq), which is a high-performance asynchronous messaging library, to enable high-performance communication between workers and PS servers. However, the vanilla ZeroMQ is based on the kernel stack. Thus, we use lwIP to replace its kernel stack via replacing the interfaces invoked in ZeroMQ. 
\section{Evaluation}\label{sec:evaluation}

\subsection{Methodology}
\noindent
\textbf{Testbed Setup.} Our testbed includes 8 physical machines connecting to an Intel Tofino programming switch (3.2T/s) that is equipped with two ARM cards (ARMv8.1 SoC, 24 cores). Each physical machine is equipped with 2 32-core Intel\textregistered Xeon\textregistered 4214R CPU, 128GB memory and Mellanox CX-5 dual-port 100G NICs; the OS is Ubuntu 16.04.

\noindent
\textbf{Benchmarks.} Our benchmark models include two types of sources: two industrial sparse DL applications and several real-world sparse models. The two industrial applications are search engine (SE) and on-line advertising (OA) from a large Internet enterprise (see Figure~\ref{fig:hotpar} for their characteristics). The real-world sparse models include DeepLight~\cite{deng2021deeplight}, LSTM~\cite{jozefowicz2016exploring} and NCF~\cite{he2017neural}, which are trained by a number of open-source datasets~\cite{clickdata,chelba2013one,harper2015movielens}.

\noindent
\textbf{Baselines.} We compare Libra with SwitchML~\cite{sapio2019scaling}, a state-of-the-art in-network aggregation solution, and the PS-lite framework\footnote{We choose PS-lite as one baseline because it is the basis of many PS-based learning systems (\eg MXNet).} which transmits non-zero gradients in form of <key, value>. To enable SwitchML to support sparse DL training, the gradients of the entire sparse DL model are transmitted in each iteration, rather than only the non-zero gradients. It is noteworthy that the default available memory of programmable switches for aggregation is limited to 1MB (5\%*20MB) for both Libra and SwitchML.

We evaluate Libra from five aspects: \one the effectiveness of sparse gradient aggregation; \two the feasibility of our sampling-based method for identifying hot model parameters; \three the benefit of the hot parameter layout and orchestration; \four the performance of the table-lookup floating-point summation compared with the float-to-integer solution used in SwitchML; \five the overhead due to the introduced mechanisms for reliability; and \six the
resource consumption on the switch data plane.

\subsection{Aggregation Throughput}
We first evaluate the aggregation throughput of Libra, and compare it with the two baselines. To this end, we use the tensors generated by training the benchmark models to test the aggregation throughput, where each worker transmits the tensors it generated when training individual models\footnote{As in~\cite{switchml}, we do not directly compare the end-to-end training performance in order to eliminate the impacts of computation capacities on workers on training performance.}. It is noteworthy that in SwitchML, each worker has to transmit all gradients in each iteration, regardless whether the gradients are zero or not. Besides, we selected the top 30,000 parameters with the most update frequency as hot parameters for OA and SE (see Figure~\ref{fig:hotpar}), top 40,000 for DeepLight, and top 60,000 for LSTM and NFC.

Figure~\ref{fig:vs-ps-lite} shows the aggregation throughput over 5 models. Libra consistently achieves higher aggregation throughput than that of the baselines, because of its sparse-awareness to aggregate only the hot parameters on programmable switches. Indeed, SwitchML falls short in terms of aggregation throughput in supporting these deep sparse training, because of its streaming-based aggregation that has to aggregate all gradients for all parameters on switches (See \S\ref{sec:limitation}). Because of the limited memory in the data plane for gradient aggregation (1MB), the benefit of in-network aggregation in SwitchML may even be out-weighted by its extra overhead, leading to even poorer performance than the PS-lite, especially for the models with high sparsity (\eg DeepLight). Note, the PS-lite follows the typical deep sparse training where only the non-zero gradients are transmitted. We also find that the advantage of Libra becomes more significant as the number of the workers gets larger, because more workers means more opportunities of in-network aggregation. Specifically, with 32 workers, the aggregation throughput can be improved by 1.5$\sim$4$\times$ in comparison with the PS-lite-sparse.

\begin{figure}[!ht]
   \centering
   \includegraphics[scale=0.27]{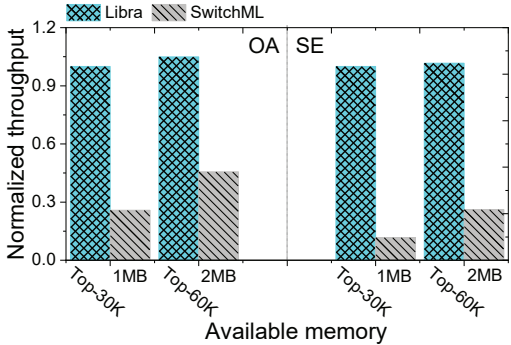}
   \caption{Aggregation throughput with different memory caps. We normalize the throughput results with that of Libra running on the default configuration.}\label{fig:memory_effect}
\end{figure}

\begin{figure}[!ht]
   \centering
   \includegraphics[scale=0.27]{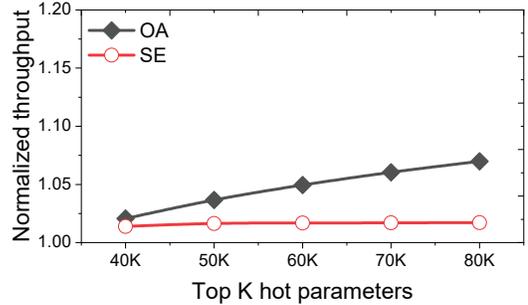}
   \caption{Libra's throughput when increasing the number of hot parameters for in-network aggregation. The throughput is normalized by that with the default configuration.}\label{fig:topk_effect}
\end{figure}

To show the impact of memory cap in the data plane on throughput, we increase the memory cap for in-network aggregation from 1MB to 2MB for SwitchML. We correspondingly double the number of hot parameters from 30K to 60K for Libra. We use 16 workers for this set of experiments. 

Figure~\ref{fig:memory_effect} shows the results. Both Libra and SwitchML benefit from increasing the switch memory cap. However, for Libra, the improvement is limited: although the offloaded hot parameters are double, the improvement is 7\% for OA and 1.7\% for SE. The reason is that, as shown in Figure~\ref{fig:topk_effect}, the extra offloaded parameters contribute to a limited amount of updates that can be aggregated.

SwitchML, on the other hand, is much sensitive to the memory cap: the throughput also doubles. This is because for SwitchML, a larger amount of on-chip memory lead to more parameters to be aggregated. Nevertheless, even with doubled memory usage, the aggregation throughput of SwitchML is only 45.6\% of the throughput of Libra with default configuration. These results show the importance of sparseness-awareness when aggregating gradients for distributed deep sparse training.

Figure~\ref{fig:memory_effect} shows the results. Both Libra and SwitchML benefit from increasing the switch memory. However, for Libra, the brought improvement is limited. Even if the offloaded hot parameters are increased up to 2.67 times, the improvement is below 7\% for OA and 1.7\% for SE. This is because that the extra offloaded parameters only correspond to a few traffic that can be aggregated. SwitchML seems more sensitive to the switch memory, and we double the used memory so that SwitchML is able to achieve more than 2.23X speedup comparing with its own default result. Indeed, for SwitchML, a large of switch registers enable SwitchML to aggregate more parameters at the same time. However, even so its aggregation throughput is far away from Libra equipped with the default configuration (only reaching 45.6\%), since SwitchML indeed aggregates too much zero gradients. In addition, the available switch memory for in-network aggregation actually is often limited, which cannot provide sufficient resources to accelerate aggregation. 

\subsection{Precision of Hot Parameter Identification}\label{sec:hot-precision}
We identify the hot parameters by running the training with a sampled small dataset (\S\ref{sec:hotidentify}). In this set of experiments, we evaluate the precision in identifying the hot parameters. To this end, we first train the sparse models on our testbed with the whole datasets, and record the update frequency of each model parameter at the PS server side. By doing so, we get the global hot parameters. Specifically, we sort the model parameters in descending order according to the update frequency, and add each time 1,000 parameters to the hot parameter list $H_g$ from the highest rank to the lowest, till the increase of the cumulative update frequency falls below a pre-defined threshold (1\% in our experiment). We then train the sparse model with the sampled datasets, and use the same method to get the hot parameter list $H_s$. Finally, we use $\lvert H_{g}\cap H_{s}\rvert$/$\lvert H_{g}\rvert$ to evaluate the precision.

\begin{figure}[!ht]
   \centering
   \includegraphics[scale=0.28]{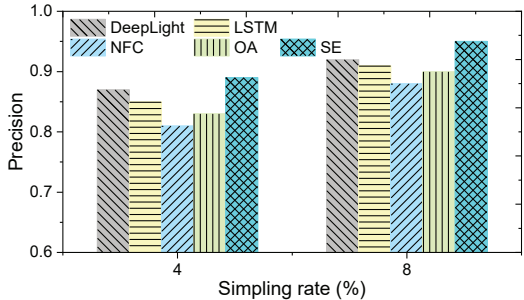}
   \caption{Precision of the sampling-based method in identifying hot parameters under different sampling rates.}\label{fig:simple}
\end{figure}

Figure~\ref{fig:simple} shows the results. For the considered benchmark models, the precision exceeds 80\% even with a sampling rate as low as 4\%. Increasing the sampling rate to 8\% improves the precision to over 90\%. These results demonstrate the high accuracy of identifying hot parameters using our sampling-based method.

\subsection{Evaluation of Hot Parameter Layout and Orchestration}\label{sec:exp_aggr}

We next evaluate the benefit of the parameter orchestration (see \S\ref{sec:para_orche}) in terms of the reduction of recirculations. In this set of experiments, we use the two benchmark models (OA and SE) from industry. We first design the parameter layout in registers and generate the packets accordingly at workers. We then count the number of recirculation operations needed on the switch and report the average number of recirculations for each packet.  For the comparative purpose, we take a random parameter placement as the baseline.

\begin{figure}[!ht]
\centering
\begin{minipage}{0.8\linewidth}
  \centering
  \includegraphics[width=\linewidth]{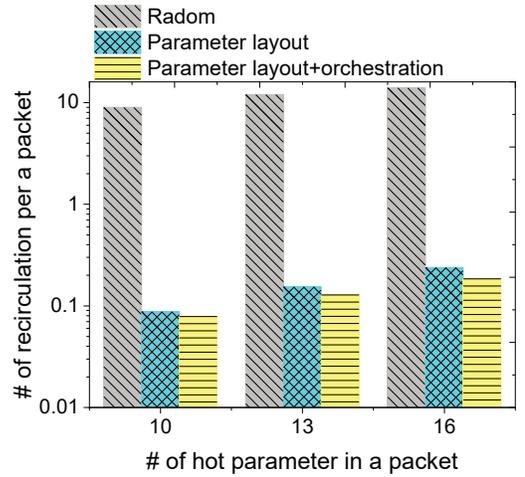}
  \centerline{(a) SE}
\end{minipage}
\begin{minipage}{0.8\linewidth}
  \centering
  \includegraphics[width=\linewidth]{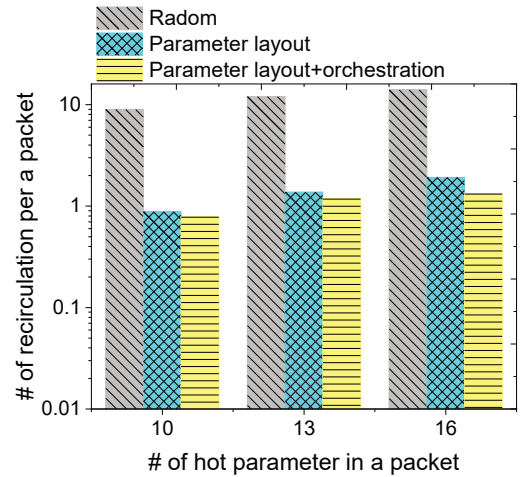}
  \centerline{(b) OA}
\end{minipage}
\caption{The number of recirculations per packet for the two benchmark models under different parameter orchestrations. Note that the $y$-axis is in log scale.}\label{fig:layout}
\end{figure}

Figure~\ref{fig:layout} shows the results, which demonstrate the effectiveness of our algorithm. Specifically, the average number of recirculations per packet is less than 1 for both models. In comparison, the random layout requires 10 recirculations per packet. The results also show the additional benefits of our gradient packaging method. We also notice that more recirculations are required in the OA model than the SE model. This is because the heat (\ie update frequency) distribution of hot parameters in OA is less biased than SE (see Figure~\ref{fig:hotpar}).

\subsection{Benefit of Floating-Point Summation}\label{sec:fp}
\begin{figure}[!ht]
   \centering
   \includegraphics[scale=0.28]{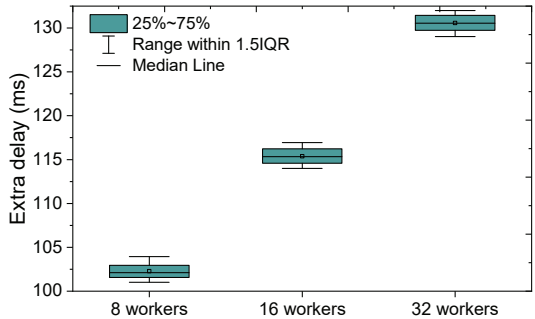}
   \caption{The extra delay incurred in each iteration for negotiating the scaling factor in the float-to-integer mechanism in SwitchML~\cite{switchml}.}
   \label{fig:negotiation_overhead}
\end{figure}

We next evaluate the benefit of the on-the-fly floating point summation enabled by our table-lookup mechanism (\S\ref{sec:approximate}). To this end, we first show in Figure~\ref{fig:negotiation_overhead} the extra delay introduced by the float-to-integer solution that is used in SwitchML~\cite{switchml} due to the scaling factor negotiation. We can see that for each iteration of negotiation, the extra delay is over 100ms when only 8 workers are used; the extra delay increases to 130ms when 32 workers are used for training. Note that our table-lookup solution does not need this negotiation and thus can save the extra delay. 

Next we evaluate the precision of the two solutions using two datasets: 1) gradients from our benchmark models ($R_1$); and 2) randomly generated floating-point numbers between (-1, 1) ($R_2$). For the float-to-integer solution, we use a negotiated scaling factor as in~\cite{switchml}; in our table-lookup mechanism, each table is equipped with 30,000 entries (large enough to hold all possible numbers in our experiments). We selected 100,000 pairs of floating-point numbers from each of the two datasets and compute the summation of each pair of floats using the two mechanisms respectively. 

\begin{table}[ht]
\centering
\caption{The precision of float summation. }
\begin{tabular}{|c|c|c|c|}
\hline
Solution& Dataset & Median &  Average\\
\hline
\multirow{2}*{\emph{Float-to-integer~\cite{switchml}}} & $R_1$ & 99.97\% & 99.89\% \\
~ & $R_2$ & 36.79\% & 62.46\%\\
\hline
\multirow{2}*{\emph{Table-lookup}} & $R_1$ & 99.92\% & 99.87\% \\
~ & $R_2$ & 100\% & 99.84\% \\
\hline
\end{tabular}\label{tab:loss}
\end{table} 

Table~\ref{tab:loss} shows the results for the two solutions. Both solutions can achieve over 99.98\% precision on the the DL benchmark dataset ($R_1$). However, on the random dataset $R_2$, the float-to-integer solution fail to provide a high precision, while the table-lookup solution still achieves a precision over 99.8\%. The reason is that the range of decimals in deep learning applications are relatively small, so the float-to-integer mechanism can find out an appropriate scaling factor. In the random decimal dataset, however, the range becomes much larger and it cannot identify a proper scaling factor. It is worth noting that, as computed in \S\ref{sec:approximate}, the table-lookup mechanism uses only about 408KB on-chip memory in the switch data plane.

In summary, the proposed table-lookup mechanism eliminates the need of scaling factor negotiation (and thus the extra delay) and achieves a high precision in gradient summation with very low overhead.

\subsection{Evaluation of Packet Loss Recovery}\label{sec:overhead}
\begin{figure}[!ht]
   \centering
   \includegraphics[scale=0.45]{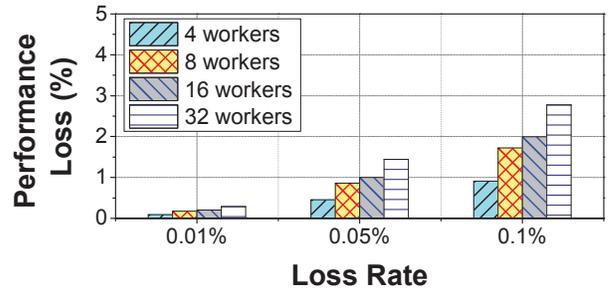}
   \caption{\small{Performance loss under different packet loss rates.}}\label{fig:packet_loss}
\end{figure}

To enable packet loss recover, the switch local CPUs are interleaved in the gradient packet processing (See \S\ref{sec:realiablity}). This may introduce extra delay. To evaluate this overhead, we trained the OA model in our testbed with packet loss rate varying from 0.01\% to 0.1\% as in~\cite{10.1145/2785956.2787496}. Figure~\ref{fig:packet_loss} shows performance loss with different packet loss rates, where the performance loss is measured by the increase of training time. We find that our packet loss recovery mechanism is practical, as even with 0.1\% packet loss rate, the performance loss is under 3\%. 

\subsection{Resource consumption on Switches}
Finally, we evaluate the resource consumption on the switch data plane. Libra uses extra switch on-chip memory for gradient caching and float-pointing operation, and consumes the pipeline stages on switches. We deploy the OA benchmark model in Libra, and use P4i, a visualization tool offered by Intel, to observe the resource consumption.

Libra consumes 9 pipeline stages in total. 
In case that Libra affects the basic functionality of a switch (\eg packet forwarding), we can utilize one recirculation to halve the the number of pipeline stages in use. In addition, Libra requires about 118KB for 30,000 hot parameters, 408.5KB for floating-point calculation,  
and 130KB for logic control. In total, it uses 656.5KB on-chip memory---only 3.21\% of the 20MB on-chip memory in our Tofino switch. Other types of resources include VLIW instructions (37 out of 384) and hash dist units (16 out of 32). 
\section{Discussion}\label{sec:discussion}
\textbf{Multi-tenant training.} 
Libra can be easily extended to multi-tenant training. Specifically, we can divide the switch on-chip memory into equal-sized parts, each of which is assigned to one training task; workers and PS servers belonging to one task label their transmitted packets with a specific job ID that will be parsed by switches. Multi-tenant training tasks can share the tables used by the table-lookup mechanism for floating-point operations. 



\noindent
\textbf{Multi-rack switches.} Note that Libra can also work with multi-rack switches in data centers. Specifically, we can designate the ToR (Top-of-rack) switches with which individual PS servers connect to run Libra's two components ($Libra_p$ and $Libra_s$) for sparse gradient aggregation. 




\noindent
\textbf{RDMA and Libra.} AllReduce implementations make use of RDMA to speed up communication. The authors in~\cite{wu2021atp} showed that in-network aggregation achieves better performance than the allreduce approach over RDMA, when training deep models for dense-data applications. We leave the comparison between Libra and RDMA, as well as the possibility of integrating them as one of our future works. 

\noindent
\textbf{Encrypted traffic.} Although AES encryption/decryption algorithms can be deployed on programmable switches directly~\cite{chen2020implementing}, Libra does not consider it necessary to accommodate for encryption traffic in this paper. This is because implementing AES encryption/decryption on programmable switches will consume not a few resources (e.g., stages and on-chip memory); Libra cannot support both sparse gradient aggregation and traffic encryption/decryption on one switch. Another possible solution is to utilize multiple switches to mitigate the limited resources, and we leave it as our further work.


\section{Related Work}\label{sec:related}

\noindent
\textbf{Distributed deep learning training.} Distributed DL leverages a cluster of nodes, each equipped with one or more GPUs, to perform sophisticated training tasks cooperatively. Overall, it either adopts centralized parameter servers~\cite{li2014scaling} or utilizes the AllReduce based training~\cite{facebook}. In addition, data parallelism~\cite{tiresias} and model parallelism~\cite{narayanan2019pipedream} have widely been adopted in distributed DL training. With this basis, many popular distributed training frameworks, such as TensorFlow~\cite{abadi2016tensorflow}, PyTorch~\cite{paszke2019pytorch} and MXNet~\cite{chen2015mxnet}, have been proposed. In this paper, Libra accelerates distributed DL training in the parameter server architecture with data parallelism.

\noindent
\textbf{In-network computation.} With the rise of SmartNICs and programmable switch-ASICs (e.g. Tofino~\cite{tofino}), in-network computation emerges~\cite{tokusashi2019case} for speed up data transmission for many applications. For example, NetCache~\cite{jin2017netcache} implements a key-value cache on programmable switches to process more than 2B queries/second; Beamer utilizes programmable switches to improve the performance of load balancers; Sailfish~\cite{pan2021sailfish} proposes a cloud-scale gateway accelerated by programmable switches so that it can process dozen of Tbps traffic; AccelTCP~\cite{moon2020acceltcp} offloads complex TCP operations to SmartNICs in order to significantly improve the host stack; Jaqen~\cite{liu2021jaqen} designs a switch-native approach for volumetric DDoS defense that can handle large-scale hybrid and dynamic attacks within seconds. Besides, in-network computation has also been recently used to accelerate distributed deep learning training via in-network gradient aggregation on programmable switches~\cite{sapio2017network,sapio2019scaling,wu2021atp,gebara2021network}. However, these approaches target dense models and they fall short when training sparse models. Thus, this paper presents Libra whose target is to accelerate sparse model training.

\noindent
\textbf{Acceleration of distributed sparse DL training.} 
NCCL~\cite{nccl}, MPI~\cite{gabriel2004open} and Gloo~\cite{gloo} design high-performance collective communication libraries that are used to accelerate distributed DL training; RDMA~\cite{guo2016rdma} is adopted to accelerate data transmission with an extra in-network support (\eg infiniband network). Other solutions reduce data transmission volume via gradient quantization or parameter synchronization. TernGrad~\cite{wen2017terngrad} quantizes floating-point gradients into three numerical levels \{-1,0,1\}; Google proposes to shorten the width of each gradient to a 4-bit vector~\cite{suresh2017distributed}. The deep model training can also be accelerated either through flow scheduling to minimize flow completion time~\cite{mai2015optimizing, chowdhury2014efficient}, or through  communication scheduling to decouple the dependence between gradients and change their transmission order~\cite{peng2019generic}~\cite{hashemi2018tictac}. These works accelerate the training at end host side, and are complementary to Libra. Recently, Omnireduce~\cite{fei2021efficient} exploits the data sparsity to improve the effective bandwidth use for distributed deep sparse training; it does not consider the highly skewed distribution of update frequency of individual parameters. Nevertheless, incorporating Libra with Omnireduce is worth further investigation. 

\section{Conclusion}\label{sec:conclusion}
In this paper, we present the design of Libra that perform in-network gradient aggregation in order to accelerate distributed sparse DL training. Overall, Libra is motivated by the key observation on the highly skewed update frequencies of parameters from industrial sparse DL models---the \emph{hot-cold} phenomenon. Based on this observation, Libra offloads the aggregation of hot parameters from PS servers to programmable switches. To this end, we carefully design the solutions that include the sample-based hot parameter identification, the parameter orchestration at both switches and workers, the table-lookup mechanism to implement the on-the-fly 32-bit floating-point operations, and also two enhancements to improve the system reliability. We implemented Libra on Intel Tofino switches and integrated it with PS-lite. Extensive experiments with different sparse models have shown the superior performance of Libra. We believe our Libra paves the way towards the high-throughput distributed sparse DL system. 





\bibliographystyle{ACM-Reference-Format}
\bibliography{reference}

\end{document}